\pdfoutput=1

\documentclass[11pt]{article}

\usepackage{ACL2023}

\usepackage{times}
\usepackage{latexsym}
\usepackage{graphicx}
\usepackage{amsmath,amsfonts}
\usepackage{listings}
\usepackage{tabularx,booktabs}
\usepackage{multirow}
\usepackage{booktabs}
\usepackage{mathtools}
\usepackage{tikz}
\usepackage{xcolor}

\usepackage[T1]{fontenc}

\usepackage[utf8]{inputenc}

\usepackage{microtype}

\usepackage{inconsolata}

\title{Code Execution with Pre-trained Language Models}

\author{
Chenxiao Liu$^{1}$\thanks{\ \ Work done during internship at Microsoft. Shuai Lu and Nan Duan are corresponding authors.},
Shuai Lu$^{2}$,
Weizhu Chen$^{2}$,
Daxin Jiang$^{2}$, \\
{\bf Alexey Svyatkovskiy$^{2}$,
Shengyu Fu$^{2}$,
Neel Sundaresan$^{2}$,
Nan Duan$^{2}$} \\
$^{1}$ Peking University \quad
$^{2}$ Microsoft \\
{\texttt jadecxliu@gmail.com}\\
{\texttt \{shuailu, wzchen, djiang\}@microsoft.com} \\
{\texttt \{alsvyatk, shengyfu, neels, nanduan\}@microsoft.com}
}

\begin{document}
\maketitle
\begin{abstract}
Code execution is a fundamental aspect of programming language semantics that reflects the exact behavior of the code. However, most pre-trained models for code intelligence ignore the execution trace and only rely on source code and syntactic structures. In this paper, we investigate how well pre-trained models can understand and perform code execution. We develop a mutation-based data augmentation technique to create a large-scale and realistic Python dataset and task for code execution, which challenges existing models such as Codex. We then present CodeExecutor, a Transformer model that leverages code execution pre-training and curriculum learning to enhance its semantic comprehension. We evaluate CodeExecutor on code execution and show its promising performance and limitations. We also demonstrate its potential benefits for code intelligence tasks such as zero-shot code-to-code search and text-to-code generation. Our analysis provides insights into the learning and generalization abilities of pre-trained models for code execution.

\end{abstract}

\section{Introduction}
\begin{figure*}[t]
  \centering  
  \includegraphics[width=0.9\linewidth]
  {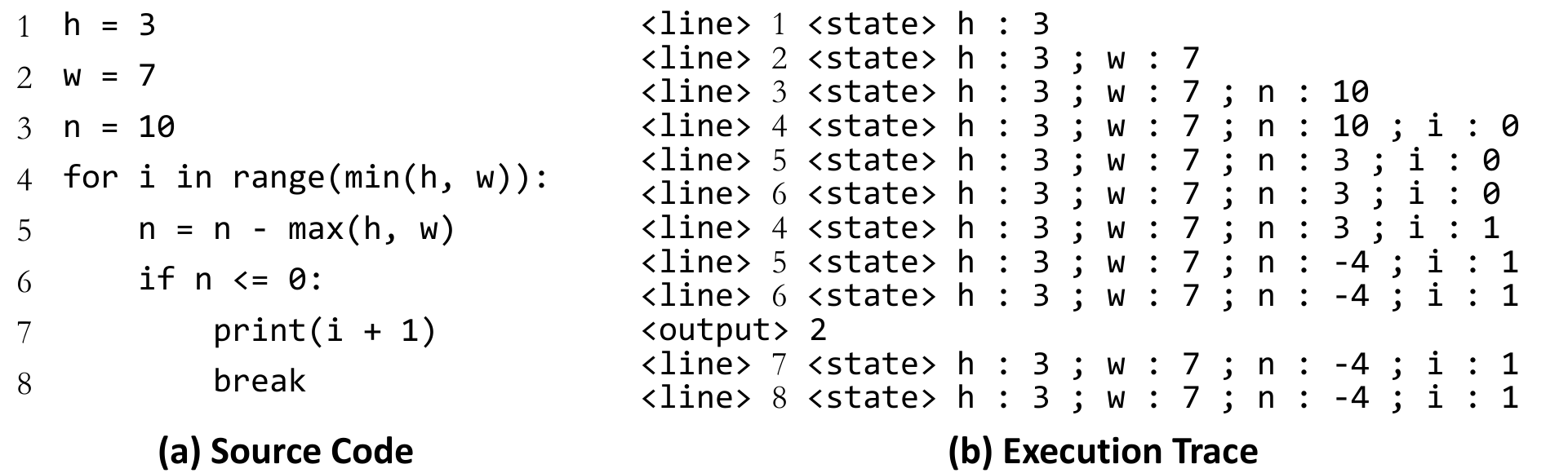} 
  \caption{Sample source code and its execution trace in the code execution task.}
  \label{fig:intro}
\end{figure*}

Pre-trained models have achieved remarkable results in natural language (NL) tasks \cite{gpt, bert, t5}, inspiring the development of pre-trained models for programming language (PL) tasks \cite{cubert, codebert, gpt-c, codet5, graphcodebert, unixcoder}. These models leverage source code and code structures, such as abstract syntax tree (AST) \cite{syncobert, unixcoder} and data flow \cite{graphcodebert}, to learn code-related tasks. These structures, while useful, are not sufficient to represent the dynamic behavior of code during execution, which is reflected in the execution trace. Using  Figure~\ref{fig:intro} as an example, the execution trace shows how code behaves during execution, reflecting the control flow and the state changes of variables. 
On the other hand, as stated by \citet{dualchannel1}, source code contains two channels of information: natural \& formal. The natural channel \cite{naturalness}, such as identifiers and comments, enables language models to be leveraged to understand code-related tasks. 
The formal channel is used by interpreters and compilers to specify execution and has precise semantics. The formal channel is unique to code and is what makes it executable. Execution trace falls into the second category since it reveals the formal channel of information that distinguishes code from natural language, as well as enabling code execution precisely \cite{dualchannel1, dualchannel2}.

In this work, we aim to teach pre-trained models the real-world code execution process. We propose CodeExecutor, a Transformer-based model that learns to execute arbitrary programs and predict their execution traces. To support pre-training on large-scale data, we construct the Python CodeNetMut dataset by producing mutations based on submissions to competitive programming problems from CodeNet \cite{codenet}, along with single-line Python transformations and programs adapted from Python official tutorial. We design a pre-training task that predicts both the line order and the intermediate states of the execution trace, and apply curriculum learning to gradually increase the difficulty of the programs.

We evaluate CodeExecutor on code execution tasks and show that it outperforms existing models and demonstrates promising capabilities. We also conduct an in-depth analysis of the model's performance and reveal its strengths and weaknesses. Furthermore, we show that CodeExecutor can improve downstream tasks like zero-shot code-to-code search and text-to-code generation, indicating the potential of leveraging execution trace to enhance code intelligence. Our models and datasets are publicly available\footnote{\url{https://github.com/microsoft/CodeBERT/tree/master/CodeExecutor}}.
In summary, the contributions of this paper are:

\begin{itemize}
\item We present the first attempt at building a large-scale pre-training dataset for real-world code execution using a mutation-based data augmentation approach.

\item We propose a novel pre-trained model named CodeExecutor that learns to predict the execution traces using a code execution pre-training task and curriculum learning.

\item We conduct a comprehensive evaluation of CodeExecutor for code execution tasks, providing a detailed understanding of the model's performance.

\item CodeExecutor significantly improves code intelligence tasks like zero-shot code-to-code search and text-to-code generation.

\end{itemize}

\section{Related Work}

\begin{table*}[t]
\centering
\small
\begin{tabularx}{\textwidth}{p{0.5cm}p{5.5cm}X}
\hline
 & \textbf{Operator}  & \textbf{Description}\\
\hline
CRP & Constant Replacement  & Change numeric and string literals.    \\
AOD & Arithmetic Operator Deletion & Delete a unary arithmetic operator ‘+’ or ‘-’. \\
AOR & Arithmetic Operator Replacement & Replace an arithmetic operator with another one. E.g. \textit{x * y} can be mutated to \textit{x / y}. \\
ASR & Assignment Operator Replacement & Substitute an extended assignment operator with another.

\\
BCR & Break Continue Replacement & Swap keywords \textit{break} and \textit{continue} in a loop body. \\
COD & Conditional Operator Deletion &  Delete unary negation operator \textit{not} or the negation of an membership operator \textit{not in}.\\
LCR & Logical Connector Replacement &  Swap logical operators \textit{and} with \textit{or} and vice versa. \\
ROR & Relational Operator Replacement &  Substitutes relational operators. E.g. \textit{x <= y} can be mutated to \textit{x > y}.  \\
SIR & Slice Index Removal &  Delete one argument of \textit{collection[start:end:step]}.  \\
OIL & One Iteration Loop  &  Execute a loop only once by adding a \textit{break} statement.  \\
RIL & Reverse Iteration Loop  &  Change direction of loop iteration by the function \textit{reversed()}. \\
ZIL & Zero Iteration Loop  & Interrupt realization of a loop during its first iteration. \\
\hline
\end{tabularx}
\caption{\label{mutation-operators}
A set of mutation operators containing 12 operators we implement to mutate code examples. 
}
\end{table*}

\subsection{Learning to Execute}
Previous works form the \textit{learning to execute} task as a problem that reads a program and computes the program’s output.
These works leverage architectures such as recurrent neural networks \cite{learn-to-execute}, graph neural networks \cite{ipagnn, ginn} and Transformers \cite{universal-tfm, Neural-Execution-Engines, mbpp, scratchpad}.
Another related task \textit{algorithm induction} is to read a short program, such as integer addition or polynomial evaluation, and computes the output. 
Algorithm induction task \cite{induction1, induction2, induction3, induction4, induction5, universal-tfm, induction6, induction7, scratchpad} targets a particular algorithm with direct algorithm-specific supervision compared with arbitrary programs in our code execution task. 

Some emerging works also employ pre-trained models to tackle the two tasks. \citet{frozen-tfm} fine-tunes a small fraction of the weights in GPT-2 \cite{gpt2} on non-language tasks, including simple algorithm induction tasks like Bit XOR. \citet{mbpp} evaluates models pre-trained on web documents and dialog data ranging in size from 2 million to 137 billion parameters and shows that largest models are generally unable to predict the output of a program, whether few-shot or fine-tuning. \citet{scratchpad} uses a "scratchpad" to store intermediate computation steps to perform multi-step computations, improving the ability of models in \citet{mbpp}.

Different from previous works that predict program’s output and mainly deal with specific algorithms, we predict the program’s whole execution trace and focus on imitating the real-world arbitrary program execution behavior. Besides, by using execution to capture code semantics, our work is beneficial for tasks related to code intelligence.

\subsection{Mathematical Problem Solving}
Mathematical problem solving is a related domain of code execution.
Recent works show the ability of language models to solve math problems, which requires learning to execute a soft algorithm to arrive at a deterministic answer.
\citet{mathqa, aqua} map math problems to operation programs and focus on sequence-to-program generation.
\citet{deepmind-math} introduce the DeepMind Mathematics dataset, which contains plug-and-chug problems such as addition, list sorting, and function evaluation. 
\citet{scaling-math} shows that the majority of problems in the DeepMind Mathematics dataset
can be straightforwardly solved with large Transformers.
\citet{math-dataset} introduces the MATH dataset, consisting of competition math problems with step-by-step solutions written in \LaTeX{} and natural languages. 
\citet{gsm8k} releases GSM8K, including grade school math questions and natural language solutions. 
Recently, \citet{algo-reason} proposes algorithmic prompting to improve the performance of large language models on math problem solving, which starts from learning skills containing addition, subtraction, multiplication, and parity.

Code execution involves calculations such as addition, subtraction, multiplication, division, exponentiation, and modulus, which are similar to solving math problems. 
With the added complexity of managing variables, data structures, control flows, and other programming concepts, learning code execution requires a different set of skills and knowledge from learning mathematics, although some overlap exists.

\section{Mutation-based Data Augmentation}\label{sec:data-aug}
The goal of \textit{code execution} task is to learn to emulate the execution without running a program by an interpreter. 
We treat the task as a generation task: given a source code $c$, the execution trace $t$ is required to be generated.
Execution trace consists of two components: one is the order in which the computer executes statements, and the other is how the states of the variables change when jumping from one statement to another.
Normally, the statements inside a program are not executed sequentially, especially in a real-world scenario where programs embody complex logic and rich semantics.
Moreover, variables relate to various types of data structures with diverse characteristics and operations.
Given the complexity and difficulty of this task, it is of great importance to build a large-scale dataset and explore the capabilities and boundaries of large language models for code execution.

\subsection{Mutating Source Code}
Constructing a large-scale Python dataset for
real-world code execution is very challenging. 
Programs retrieved from software development platforms such as GitHub \footnote{\url{https://github.com/}} are mostly not executable at scale, as they depend on specific external resources which are not easily available. Examples of external resources include program inputs, file contents, external modules, and third-party packages. For the same reason, it is not practical to collect programs from posts in coding question-answering websites like StackOverflow \footnote{\url{https://stackoverflow.com/}}.

We build the Python code execution dataset based on submissions to competitive programming problems from CodeNet benchmark \cite{codenet}. We run each submission in a sandbox environment to get the execution trace and  filter out programs that exceed time and trace limits or result in runtime errors.

To construct a large-scale dataset of executable programs, 
we propose a mutation-based data augmentation approach. 
For each submission, the approach modifies some parts of a program to generate diverse mutants, leading to different execution traces. 
Specifications of these modifications are called mutation operators. 
It is inspired by \textit{mutation testing} \cite{mutation-test1, mutation-test2} in software engineering, a popular technique that supports the design of high-quality test suites for programs.
Following \citet{derezinska2014operators} that applies mutation testing technique to Python programs, we first present a set of mutation operators as shown in Table \ref{mutation-operators}.
Most of them correspond to selected operators used in strongly typed general purpose languages and are adopted to the Python language. 
Operators designed for Python features are also included, such as Slice Index Removal (SIR) and Reverse Iteration Loop (RIL). 
Then we convert a program into an AST and extract its node type information to get a candidate list of all mutable literals, operators and statements. 
Finally, we generate mutants and eliminate those that are not executable.
We use the CodeNet Mutants (CodeNetMut) to build the pre-training dataset. 
Greater detail of the dataset generation process can be found in Appendix \ref{sec:appendix-dataset}.

\subsection{Dataset Construction}
Given the difficulty of training the model on real-world complete programs, we build two simpler datasets along with CodeNetMut for pre-training. 

\begin{figure}[t]
  \centering  
  \includegraphics[width=\linewidth]
  {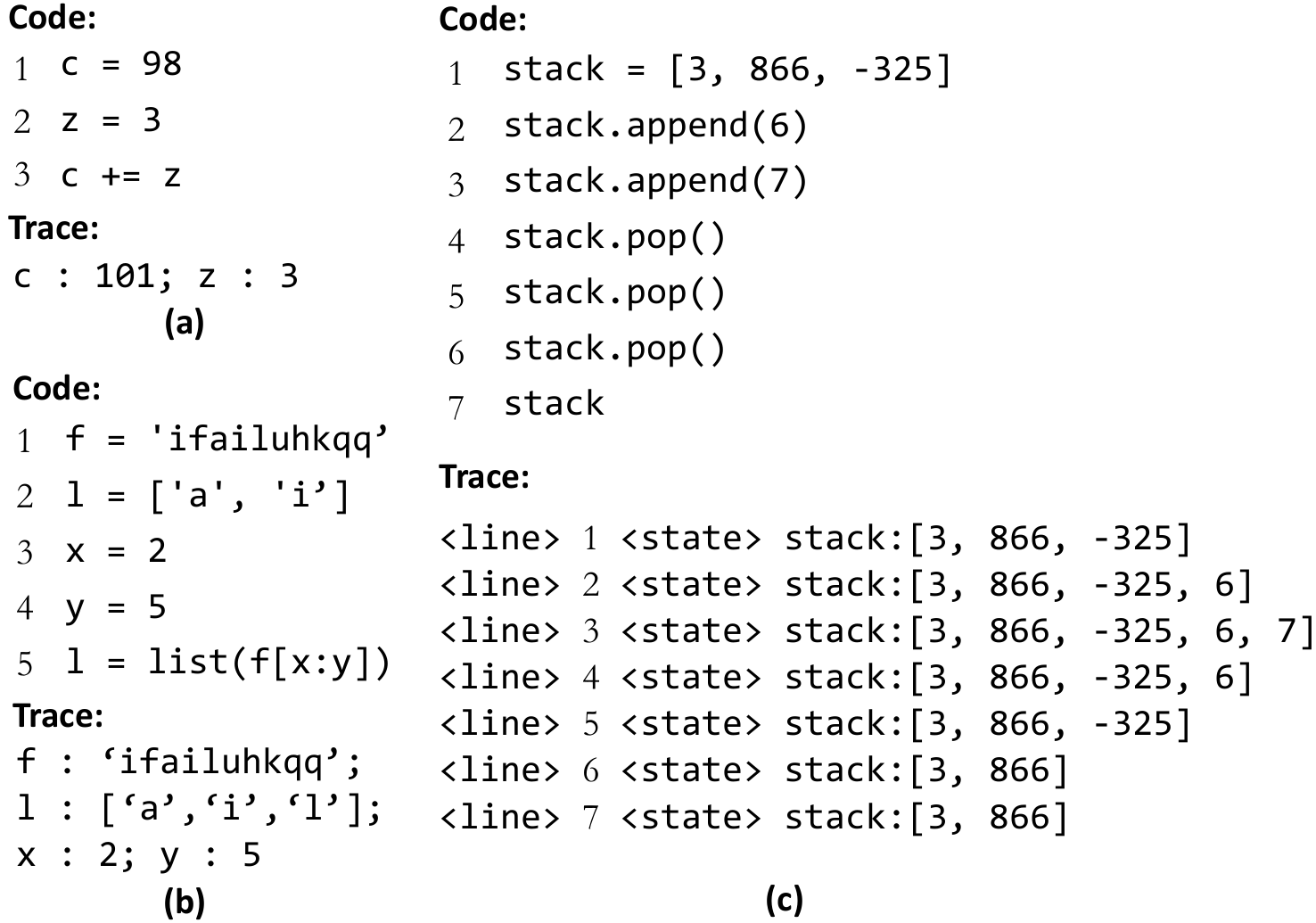} 
  \caption{(a) and (b) are examples from the SingleLine dataset. (c) is an example from the Tutorial dataset.}
  \label{fig:example-singleline-tutorial}
\end{figure}

The first is the Python SingleLine dataset collected by Fraser Greenlee \footnote{\url{https://www.kaggle.com/frasergreenlee/python-state-changes}}, which consists of nearly nine million examples of single-line transformations. Each example contains several variables specified in initial values, a single line of Python code, and the new set of variables and values resulting from executing that line.
We combine the first two as the input code, and use the last one as the target trace. We do not re-execute the dataset. When pre-training on SingleLine data, we only ask the model to predict the final states of the last code line without line-by-line illustration. 
Figure \ref{fig:example-singleline-tutorial} (a)(b) show examples of these data.
Since individual lines of code constitute real-world complex programs, the dataset serves as a foundation for learning about code execution.

The second is the Python Tutorial dataset. This dataset is created by crawling and filtering all the executable code examples that appear in the official Python tutorial \footnote{\url{https://docs.python.org/3/tutorial}}.
The official tutorial introduces the basic concepts and most noteworthy features of the Python language.
To generate this dataset, we apply the Constant Replacement operator (first row in Table \ref{mutation-operators}) to change numeric literals into diverse values. This approach results in 3.4 million programs.
Figure \ref{fig:example-singleline-tutorial} (c) shows an example of a mutant.
While the Tutorial dataset is not comprehensive and does not cover every single feature, it provides a good representation of Python’s flavor and style, which offers valuable supervision for modeling the execution of commonly used code blocks.

Therefore, the Python Code Execution datasets are a series of datasets following an easy-to-hard paradigm, including the SingleLine dataset, Tutorial dataset, and CodeNetMut dataset.

\section{CodeExecutor}
Our CodeExecutor utilizes a Transformer-based framework to learn code execution through pre-training.
We will first describe the model architecture (§\ref{sssec:4.1}), then the pre-training task (§\ref{sssec:4.2}), and finally, the curriculum learning strategy (§\ref{sssec:4.3}).

\subsection{Model Architecture}\label{sssec:4.1}
 The model is based on Transformer and adopts the same architecture as UniXcoder \cite{unixcoder}. UniXcoder is a unified cross-modal pre-trained model for programming language which has encoder-only, decoder-only and encoder-decoder modes. 
It utilizes mask attention matrices
\cite{unified-lm} with prefix adapters to control
the behavior.
We take the encoder-decoder manner by using a special token $[E2D]$ as the prefix in front of the input. 
CodeExecutor consists of 12 Transformer layers.
Each transformer layer is architecturally
identical, containing a multi-headed self-attention pooling \cite{vaswani2017attention} followed
by a feed forward network.

\subsection{Pre-training Task}\label{sssec:4.2}
We propose a new pre-training task called code execution.
Our motivation for the task is to improve the ability of our model to understand and execute code. 
Traditional pre-training tasks such as language modeling or denoising objective do not involve code execution, and thus, models trained on these tasks have limited ability to execute code. 
By pre-training our model on the task of code execution, we aim to improve its ability by learning useful patterns from bimodal data of code and trace. 
This will enable our model to generate more accurate traces and understand the behavior of the code, which is crucial for a wide range of code intelligence applications that require code understanding. With the knowledge of how the code works, the model can better understand the underlying logic of the code and use that understanding to better perform these tasks.

We continue pre-training UniXcoder on the task.
At the pre-training stage, our model receives code as inputs and learns to generate traces.
To facilitate a better understanding of code, special tokens $[i]$ indicating line numbers and $[INDENT]$ $[DETENT]$ indicating indentation are inserted into the code. Each line in trace can be represented as $[LINE], [i], [STATE], v_1, :, s_1, [DICTSEP], ...,[DICTSEP], v_k, :, s_k,$ $[STATEEND]$, where $k$ denotes the number of variables and the state of $k$-th variable $v_k$ is $s_k$. 
The symbol $[DICTSEP]$ separates the pairs within the dictionary and $[STATEEND]$ indicates the end of the states.
This representation allows our model to learn the state of variables at each step of the execution, which is crucial for understanding the behavior of the code.

\subsection{Curriculum Learning}\label{sssec:4.3}
To improve the generalization capacity, we follow the curriculum learning strategy during pre-training.
Curriculum learning \cite{cl} (CL) is a learning strategy that starts from easy instances and then gradually handles harder ones, which imitates the meaningful learning order in human curricula. 
In our pre-training process, we organize the learning of the Python code execution datasets according to a curriculum that starts with simple instances, i.e. SingleLine data.
First, we employ all the 9 million SingleLine transformations to pre-train CodeExecutor until convergence.
To achieve a balanced dataset, we then reserve 3 million instances in SingleLine that are most difficult for our model to generate and add Tutorial data into the pre-training corpus. 
We further add CodeNetMut data into the pre-training corpus and pre-train the model to converge on all the examples.
To help distinguish difficulty level, we add a prefix $p\in \{[SINGLELINE],$ $[TUTORIAL],[CODENETMUT]\}$ in front of the input, indicating the kind of data, e.g. $[SINGLELINE]$ means receiving SingleLine data. 
More details about pre-training settings and model configurations can be found in Appendix \ref{sec:appendix-model}.

\section{Experimental Setup}
\begin{table}
\centering
\small
\resizebox{\linewidth}{!}{
\begin{tabular}{lccc}
\toprule
 & \textbf{SingleLine} & \textbf{Tutorial} & \textbf{CodeNetMut} \\
Difficulty Level & Easy & Medium & Hard \\
\midrule 
Language & Python & Python & Python \\
Pre-train \# & 8,950,959 & 3,422,943 & 2,838,644 \\
Test \# & 7,968 & 13,744 & 19,541 \\
Avg Code Len & 3.28 & 4.90 & 8.26\\
Avg Trace Len & 1.00 & 11.89 & 22.80\\
Avg State Num & 2.44 & 1.34 & 3.67 \\
\bottomrule
\end{tabular}
}
\caption{\label{pretrain-data}
Statistics of pre-training dataset. ``Avg Code Len'' and ``Avg Trace Len'' represent the average number of lines in a program and a trace, respectively. ``Avg State Num'' denotes the average of the maximum number of states reached per line in a trace.
}
\end{table}

\subsection{Dataset} 
We build our pre-training dataset as described in Section \ref{sec:data-aug}. Table \ref{pretrain-data} shows some basic statistics. 
The 19,541 examples in CodeNetMut test split are from 39 unseen programming problems in CodeNet and have not undergone the mutation process. 
Additionally, we held out 10k programs from each dataset as a validation split during pre-training.
For Tutorial and CodeNetMut, the ground truth trace is the execution result of the whole program.
For SingleLine, since the instances are simple programs consisting of variable declarations and one-line transformations, the model is only asked to predict the final states of variables, which is presented in the form of a one-line trace.
We observe the average length of code and trace in CodeNetMut are about twice as long as those in Tutorial. 
Also, executing programs in CodeNetMut requires managing a larger number of variables in varying states.

\begin{table*}
\centering
\small
\begin{tabular}{llcccccccc}
\toprule
\multirow{2}{*}{\textbf{Dataset}} & \multirow{2}{*}{\textbf{Model}} & \multicolumn{2}{c}{\textbf{General}} & \multicolumn{3}{c}{\textbf{Line}} & \multicolumn{3}{c}{\textbf{Identifier}} \\
\cmidrule(r){3-4} \cmidrule(r){5-7} \cmidrule(r){8-10}
& & Output Acc. & Trace Acc. & Precision & Recall & F1 & Precision & Recall & F1 \\
\midrule
\multirow{2}{*}{SingeLine} & Codex & - & 36.87 & 36.87 & 36.87 & 36.87  & 71.87 & 69.34 & 70.58 \\
& CEL-S1 & - & 93.32 & 93.32 & 93.32 & 93.32 & 96.94 & 96.86 & 96.90 \\
& CodeExecutor & - & \textbf{94.03} & \textbf{94.03} & \textbf{94.03} & \textbf{94.03} & \textbf{97.28} & \textbf{97.18} & \textbf{97.23} \\
\midrule
\multirow{4}{*}{Tutorial} & Codex & 13.07 & - & - & - & - & - & - & - \\
& CEL-S2 & \textbf{79.51} & \textbf{85.59} & \textbf{95.94} & \textbf{84.24} & \textbf{89.71} & \textbf{97.29} & \textbf{87.30} & \textbf{92.02} \\
 & CEL-S3 & 7.89 & 8.35 & 26.58 & 21.33 & 23.67 & 26.36 & 19.47 & 22.40 \\
& CodeExecutor & 76.42 & 80.09 & 94.49 & 76.74 & 84.70 & 95.91 & 69.15 & 80.36 \\
\midrule
\multirow{2}{*}{CodeNetMut} & Codex & 17.45 & - & - & - & - & -  & - & -\\
& CEL-S3 & 43.80 & 29.44  & 59.32 & 41.76 & 49.01 & 68.30 & 41.69 & 51.78 \\
& CodeExecutor & \textbf{48.06} & \textbf{33.38} & 58.70 & \textbf{43.48} & \textbf{49.96} & 67.81 & \textbf{45.29} & \textbf{54.31} \\
&  \quad -w/o CL & 45.93 & 30.98 & \textbf{60.21} & 42.45 & 49.79 & 
\textbf{68.55} & 41.58 & 51.76 \\
\bottomrule
\end{tabular}
\caption{\label{auto-evaluation}
Results on the code execution task. In the Tutorial and CodeNetMut datasets, Codex cannot generate execution traces in a uniform format. Therefore, we only report the output accuracy of Codex in these datasets.
}
\end{table*}

\subsection{Models} 
We evaluate several models on code execution task. 
\textbf{Codex} model \texttt{code-cushman-001} is a specialized GPT model fine-tuned on GitHub code \cite{codex}. We use few-shot learning by giving Codex three code and execution trace pairs for the code execution task.
CodeExecutorLimited (\textbf{CEL}) is a three-stage model pre-trained with the code execution objective. 
CEL can only access limited data in each stage, as opposed to \textbf{CodeExecutor} which can utilize all the datasets simultaneously (see Appendix \ref{appedix:stages} for a detailed comparison).
It is initialized using the publicly available checkpoint of UniXcoder and continues to be trained with SingleLine data, resulting in the model CodeExecutorLimited-Stage1, which we call \textbf{CEL-S1}.
In the second stage, we initialize it with CEL-S1 and employ Tutorial data to pre-train, so we get the model \textbf{CEL-S2}.
By continuing pre-training CEL-S2, we use CodeNetMut to improve the capacity of executing real-world programs at the third stage. \textbf{CEL-S3} is produced after these stages mentioned above. 
CodeExecutor without Curriculum Learning(\textbf{CodeExecutor  w/o CL}) is a single-stage model trained on all three datasets together.

\subsection{Evaluation Metrics} 
We test model capabilities of executing code on the test sets from three datasets.
We measure functional correctness of the sampled trace from three perspectives. 
We report output accuracy and trace accuracy to evaluate the general aspect. 
\textbf{Output accuracy} checks if the model prints the same message as the code execution, calculated only for programs with standard output.
\textbf{Trace accuracy} checks if the model produces the same trace as the code execution, regardless of the order of states in a line of the trace.
To evaluate the correctness of each line and the states of identifiers in the trace, we also assess per-line score and identifier score. 
\textbf{Line precision} is determined by the ratio of correctly identified lines among all the lines in the traces generated by the model. 
\textbf{Line recall} is the ratio of correctly identified lines predicted by the model among all the lines in the ground truth traces. 
Similarly, we also calculate scores for the identifiers in the trace.

To deepen our understanding of model behavior and error modes, we also conduct a qualitative analysis by examining samples.

We randomly sample 50 code-trace pairs from the test set and ask two programmers with at least 5 years of experience to evaluate whether CodeExecutor executes a program correctly in 7 aspects.
The category \textit{Basic} includes basic knowledge for a Python beginner like math operators, augmented assignment operators, comparison operators, variables.
The category \textit{Lists, Tuples, etc.} consists of typical Python data structures, such as lists, tuples, dictionaries, sets, and related manipulation functions.
As shown in Table \ref{human-evaluation}, we build the taxonomy, along with a handbook to guide classification. 
Each reviewer examines the generated trace line by line and counts the occurrence frequency of each category.
 They count all these categories if a trace line involves multiple categories. When an error occurs, they identify which kind of knowledge category the model mistakes. Finally, they work together to discuss the divergence of error attribution and come to an agreement.

\section{Results and Analysis}
\begin{figure*}[t]
  \centering  
  \includegraphics[width=0.98\linewidth]
  {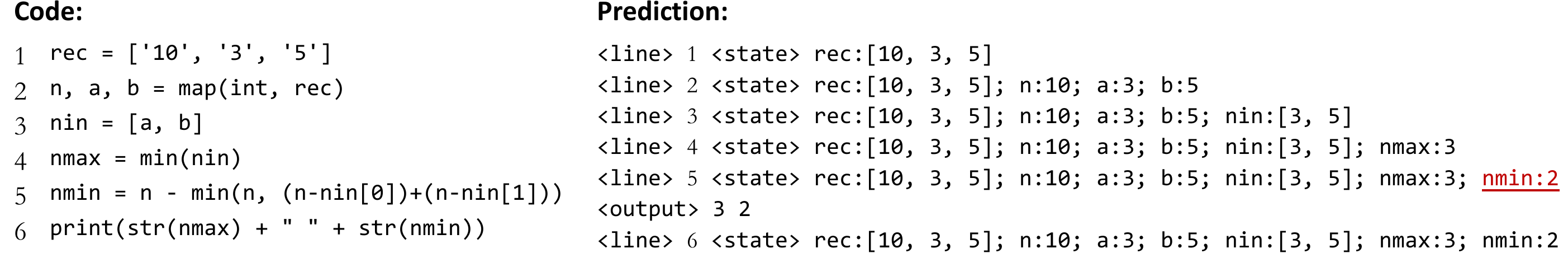} 
  \caption{An Example from CodeNetMut test split, where CodeExecutor produces an imperfect prediction, with the mistake highlighted by an underline.}
  \label{fig:example-e}
\end{figure*}

In this section, we evaluate CodeExecutor on code execution task(§\ref{sec:exp-overall}), conduct an in-depth analysis to understand model behavior and error mode (§\ref{sec:exp-study}), followed by two downstream tasks (§\ref{sec:exp-downstream}).

\subsection{Overall Results} \label{sec:exp-overall}
We evaluate the performance of models on SingleLine, Tutorial and CodeNetMut datasets.

We show the result of \textbf{SingleLine} in Table \ref{auto-evaluation} (top). 
CodeExecutor is able to execute around 94\% of single-line transformations correctly, while Codex fails to do so in most cases.
CodeExecutor also brings a 0.7\% improvement over CEL-S1, indicating learning hard programs during pre-training helps better solve easier examples.
Since each SingleLine program always produces a one-line trace without standard outputs, we do not report output accuracy, and the line precision/recall scores are equal to trace accuracy.

For the \textbf{Tutorial} experiments in Table \ref{auto-evaluation} (medium), CodeExecutor significantly outperforms Codex on output accuracy (76.42\% vs.13.07\%).
The lower score of CodeExecutor compared to CEL-S2 suggests a discrepancy between code examples in tutorials and CodeNet since the Tutorial dataset is composed of mutants from only a few programs in tutorial websites, limiting its diversity.
CEL-S3 struggles to produce traces, indicating that it forgets most knowledge acquired in Tutorial data in the last training stage.

\textbf{CodeNetMut} results are much lower than those in SingleLine and Tutorial datasets, which shows that it is more challenging to generate traces in real-world scenarios.
CodeExecutor produces the correct output for nearly half of the examples (48.06\%), and about a third of the traces are the exact match for the ground truth (33.38\%). By pre-training on the code execution task, CodeExecutor boosts the performance of output by 30.6\% absolute points over Codex. Besides, CodeExecutor yields 4.3\% output accuracy score and 3.9\% trace accuracy score improvement than CEL-S3, which indicates the effectiveness of the training strategy described in \ref{sssec:4.3}. 
After removing curriculum learning, the output accuracy score drops from 48.06\% to 45.93\% and the trace accuracy score drops from 33.38\% to 30.98\%, which shows the contribution of curriculum learning.

These results demonstrate that the code execution task is challenging for pre-trained models on source code like Codex. However, our CodeExecutor model can achieve high performance to execute simple programs and are capable of predicting complex execution traces for real-world programs.

\begin{table}
\centering
\small
\begin{tabular}{lccc}
\toprule
  \textbf{Category} & \textbf{Total} & \textbf{Correct} & \textbf{Accuracy} \\
\midrule
Basic & 204 & 183 & 89.71 \\
Built-in Functions & 42 & 35 & 83.33 \\
Lists, Tuples, etc. & 44 & 34 & 77.27 \\
Strings & 19 & 10 & 52.63 \\
Conditional Statements & 60 & 57 & 95.00 \\
Loops & 25 & 21 & 84.00 \\
Function Calls & 5 & 5 & 100.00 \\
\bottomrule
\end{tabular}
\caption{\label{human-evaluation}
Human evaluation results. To evaluate the capability of CodeExecutor, we classify Python programming knowledge into seven categories and manually analyze whether the generated trace is correct or wrong when dealing with these categories. The third category includes Python data structures, such as lists, tuples, dictionaries and sets.
}
\end{table}

\subsection{In-depth Study on Model Performance}\label{sec:exp-study}
We conduct a qualitative analysis of model performance by examining samples (Table \ref{human-evaluation}), resulting in the following findings. More examples can be found in Appendix \ref{appedix:examples}.


\paragraph{The Model Typically Has a Basic Sense of Control Flows}
Conditional statements, loops, and function calls reveal the control flow of the program.
Control flow reflects the order in which the program's code executes.
It is important for understanding a program and is often complex, as it controls the code through certain decisions and monitors which statements need to be executed and which should be skipped. 
From Table \ref{human-evaluation}, we find that CodeExecutor has a rudimentary understanding of high-level multi-line control flows, especially expert at conditional statements and function calls. 57 out of 60 conditional statements and all 5 calls to user-defined functions are predicted correctly. The accuracy of loops is 84\%, while the incorrect loops undergo wrong iterative times. 
Take Figure \ref{fig:intro} (a) as an example. CodeExecutor predicts exactly the same trace as the ground truth in (b). Our model recognizes that the for loop occurred on line 4 will execute several times. 
In the second iteration, \textit{``n''} meets the condition of \textit{``n <= 0''}, resulting in the \textit{``break''} statement and terminating the loop. The model behaves well on the code block in the for loop, showing its capacity of understanding control flows.

\paragraph{The Model Struggles to Handle the Intricacies of Operations, Particularly in Relation to Data Structures}
Complex programs often involve multiple categories of programming knowledge. 
Figure \ref{fig:example-e} shows an example that uses lists and strings. 
It determines the maximum and minimum possible number of people among \textit{``n''}, who subscribe to both Newspaper \uppercase\expandafter{\romannumeral1} and \uppercase\expandafter{\romannumeral2}, given that \textit{``a''} people subscribe to  \uppercase\expandafter{\romannumeral1} and \textit{``b''} people subscribe to \uppercase\expandafter{\romannumeral2}.
CodeExecutor incorrectly calculates \textit{``nmin''} in line 5, expected 0 but got 2.
This calculation involves retrieving values from a list, performing additions, subtractions, and using the "min" function. The compositionality of these operations makes it challenging for our model to fully comprehend the code and generate accurate states.
Additionally, as presented by the relatively low accuracy on ``Lists, Tuples, etc.'' (77.27\%) and ``Strings'' (52.63\%) in Table \ref{human-evaluation}, we observe that the model falls short of understanding data structures like lists and strings. 
The understanding of data structures requires the model to learn the behavior of objects after they are created, modified, added or deleted. These operations can be changeable and challenging for the model to grasp. 
This suggests that the model may struggle with complex programs that involve multiple operations and data structures.

\begin{table}
\centering
\small
\begin{tabular}{lc}
\toprule
\textbf{Model} & \textbf{MAP} \\
\midrule 
GraphCodeBERT &  23.08 \\
+ CodeExecutor &  \textbf{55.94} \\
\midrule 
UniXcoder & 71.86 \\
+ CodeExecutor & \textbf{79.13} \\
\bottomrule
\end{tabular}
\caption{\label{downstream-task}
MAP score (\%) on code-to-code search task in zero-shot setting.
}
\end{table}

\subsection{Downstream Tasks}\label{sec:exp-downstream}

To verify the effectiveness of CodeExecutor in representing code semantics, we apply it to two code intelligence tasks -- the zero-shot code-to-code-search task and text-to-code generation task. 
\paragraph{Zero-shot Code-to-code Search} The task is introduced by \citet{unixcoder}. To avoid duplication between the associate dataset and our pre-training corpus, we construct a new dataset by collecting 9,987 Python functions from CodeNet \cite{codenet}.
Each function solves one of the 48 problems. Given one function, we retrieve all the functions that solve the same problem.

We first use the mean vectors of last hidden states of a baseline model to calculate the similarity between two functions. 
To explore how code execution facilitates code-to-code-search, we execute each function by providing a test case. 
We then utilize the program outputs extracted from the execution trace generated by CodeExecutor, and sort the candidates according to the edit similarity compared with outputs of the query program. 

From table \ref{downstream-task}, we find that CodeExecutor boosts over 32.8 points compared with GraphCodeBERT \cite{graphcodebert}, and provides about 7.2 points improvement compared with UniXcoder, showing that code execution can significantly enhance the comprehension of code semantics.

\paragraph{Text-to-code Generation} We use HumanEval benchmark \cite{codex} which includes 164 human-written programming problems.

We first leverage Codex (\texttt{code-cushman-001}) to generate 200 solutions for each problem. Then we use CodeExecutor to predict the outputs of each solution by feeding example test cases in problem descriptions. We rank the 200 solutions by the edit similarity between their outputs and expected outputs. Finally, we evaluate the correctness of the first 50 solutions for each problem.
Note that different from other filtering strategies, our method doesn't need a real-world code executor but only uses models to predict the execution results.

\begin{table}
\centering
\small
\begin{tabular}{lcc}
\toprule
\textbf{Model} & \textbf{Pass@1} & \textbf{Pass@10} \\
\midrule 
Codex & 12.48 & 45.59 \\
+ CodeExecutor & \textbf{17.87} & \textbf{49.69} \\
\bottomrule
\end{tabular}
\caption{\label{downstream-task-generation}
Results on HumanEval benchmark for the text-to-code generation task. 50 solutions are evaluated for each problem in both settings.
}
\end{table}

Table \ref{downstream-task-generation} demonstrates that with CodeExecutor as a solution filter, the performance of text-to-code generation is improved, indicating CodeExecutor is beneficial to other code intelligence tasks.

\section{Conclusion}
We propose a mutation-based data augmentation method to create a large and realistic Python code execution dataset and task, which pose a significant challenge for current models such as Codex. We develop CodeExecutor, a Transformer model that leverages code execution as a pre-training objective and adopts a curriculum learning strategy. CodeExecutor not only outperforms existing models on code execution, but also demonstrates its generalizability to downstream tasks such as code-to-code search and text-to-code generation. Our work offers a novel and effective solution for code execution and other code intelligence tasks.

\section*{Limitations}
Several limitations of CodeExecutor, such as its application to only Python, the lack of faithfulness in the results produced, and the maximum length limit for trace generation, point toward interesting directions for future work.

\paragraph{Programming Language} One limitation of our current model is that it is currently only applied to Python, which limits its use and effectiveness in executing programs written in other programming languages. This highlights the need for future work to expand the model's applicability to other languages.

\paragraph{Faithfulness} The result may not be faithful enough when handling difficult examples, such as those with complex logic, long loops, or many branches. For example, we observe that in two complicated programs that both contain the assignment \textit{``alpha = list('abcdefg')''}, our model correctly predicts the value of \textit{``alpha''} in one case but incorrectly in the other.
The lack of faithfulness needs to be studied for further research on code execution.

\paragraph{Generation Window Size} We limit the length of generated trace to 1024 tokens. It can be a limitation for programs with long execution traces, particularly those with loops. Improving the ability of Transformers to handle longer sequences \cite{long-range, tay2022efficient}  would likely be beneficial for the code execution task.

\section*{Ethical Statement}
The work is conducted in compliance with ethical principles. The datasets introduced in this paper only used publicly available data. The annotation in human evaluation was conducted by two authors of the paper, and thus there are no associated concerns, e.g. regarding compensation. Therefore, there are no potential risks associated with the research.

\bibliography{anthology,custom}

\begin{thebibliography}{46}
\expandafter\ifx\csname natexlab\endcsname\relax\def\natexlab#1{#1}\fi

\bibitem[{Amini et~al.(2019)Amini, Gabriel, Lin, Koncel{-}Kedziorski, Choi, and
  Hajishirzi}]{mathqa}
Aida Amini, Saadia Gabriel, Shanchuan Lin, Rik Koncel{-}Kedziorski, Yejin Choi,
  and Hannaneh Hajishirzi. 2019.
\newblock \href {https://doi.org/10.18653/v1/n19-1245} {Mathqa: Towards
  interpretable math word problem solving with operation-based formalisms}.
\newblock In \emph{Proceedings of the 2019 Conference of the North American
  Chapter of the Association for Computational Linguistics: Human Language
  Technologies, {NAACL-HLT} 2019, Minneapolis, MN, USA, June 2-7, 2019, Volume
  1 (Long and Short Papers)}, pages 2357--2367. Association for Computational
  Linguistics.

\bibitem[{Austin et~al.(2021)Austin, Odena, Nye, Bosma, Michalewski, Dohan,
  Jiang, Cai, Terry, Le, and Sutton}]{mbpp}
Jacob Austin, Augustus Odena, Maxwell~I. Nye, Maarten Bosma, Henryk
  Michalewski, David Dohan, Ellen Jiang, Carrie~J. Cai, Michael Terry, Quoc~V.
  Le, and Charles Sutton. 2021.
\newblock \href {http://arxiv.org/abs/2108.07732} {Program synthesis with large
  language models}.
\newblock \emph{CoRR}, abs/2108.07732.

\bibitem[{Bengio et~al.(2009)Bengio, Louradour, Collobert, and Weston}]{cl}
Yoshua Bengio, J{\'{e}}r{\^{o}}me Louradour, Ronan Collobert, and Jason Weston.
  2009.
\newblock \href {https://doi.org/10.1145/1553374.1553380} {Curriculum
  learning}.
\newblock In \emph{Proceedings of the 26th Annual International Conference on
  Machine Learning, {ICML} 2009, Montreal, Quebec, Canada, June 14-18, 2009},
  volume 382 of \emph{{ACM} International Conference Proceeding Series}, pages
  41--48. {ACM}.

\bibitem[{Bieber et~al.(2020)Bieber, Sutton, Larochelle, and Tarlow}]{ipagnn}
David Bieber, Charles Sutton, Hugo Larochelle, and Daniel Tarlow. 2020.
\newblock \href
  {https://proceedings.neurips.cc/paper/2020/hash/62326dc7c4f7b849d6f013ba46489d6c-Abstract.html}
  {Learning to execute programs with instruction pointer attention graph neural
  networks}.
\newblock In \emph{Advances in Neural Information Processing Systems 33: Annual
  Conference on Neural Information Processing Systems 2020, NeurIPS 2020,
  December 6-12, 2020, virtual}.

\bibitem[{Casalnuovo et~al.(2020)Casalnuovo, Barr, Dash, Devanbu, and
  Morgan}]{dualchannel1}
Casey Casalnuovo, Earl~T. Barr, Santanu~Kumar Dash, Prem Devanbu, and Emily
  Morgan. 2020.
\newblock \href {https://doi.org/10.1145/3377816.3381720} {A theory of dual
  channel constraints}.
\newblock In \emph{{ICSE-NIER} 2020: 42nd International Conference on Software
  Engineering, New Ideas and Emerging Results, Seoul, South Korea, 27 June - 19
  July, 2020}, pages 25--28. {ACM}.

\bibitem[{Chakraborty et~al.(2022)Chakraborty, Ahmed, Ding, Devanbu, and
  Ray}]{dualchannel2}
Saikat Chakraborty, Toufique Ahmed, Yangruibo Ding, Premkumar~T. Devanbu, and
  Baishakhi Ray. 2022.
\newblock \href {https://doi.org/10.1145/3540250.3549162} {Natgen: generative
  pre-training by "naturalizing" source code}.
\newblock In \emph{Proceedings of the 30th {ACM} Joint European Software
  Engineering Conference and Symposium on the Foundations of Software
  Engineering, {ESEC/FSE} 2022, Singapore, Singapore, November 14-18, 2022},
  pages 18--30. {ACM}.

\bibitem[{Chen et~al.(2021)Chen, Tworek, Jun, Yuan, de~Oliveira~Pinto, Kaplan,
  Edwards, Burda, Joseph, Brockman, Ray, Puri, Krueger, Petrov, Khlaaf, Sastry,
  Mishkin, Chan, Gray, Ryder, Pavlov, Power, Kaiser, Bavarian, Winter, Tillet,
  Such, Cummings, Plappert, Chantzis, Barnes, Herbert{-}Voss, Guss, Nichol,
  Paino, Tezak, Tang, Babuschkin, Balaji, Jain, Saunders, Hesse, Carr, Leike,
  Achiam, Misra, Morikawa, Radford, Knight, Brundage, Murati, Mayer, Welinder,
  McGrew, Amodei, McCandlish, Sutskever, and Zaremba}]{codex}
Mark Chen, Jerry Tworek, Heewoo Jun, Qiming Yuan, Henrique~Ponde
  de~Oliveira~Pinto, Jared Kaplan, Harrison Edwards, Yuri Burda, Nicholas
  Joseph, Greg Brockman, Alex Ray, Raul Puri, Gretchen Krueger, Michael Petrov,
  Heidy Khlaaf, Girish Sastry, Pamela Mishkin, Brooke Chan, Scott Gray, Nick
  Ryder, Mikhail Pavlov, Alethea Power, Lukasz Kaiser, Mohammad Bavarian,
  Clemens Winter, Philippe Tillet, Felipe~Petroski Such, Dave Cummings,
  Matthias Plappert, Fotios Chantzis, Elizabeth Barnes, Ariel Herbert{-}Voss,
  William~Hebgen Guss, Alex Nichol, Alex Paino, Nikolas Tezak, Jie Tang, Igor
  Babuschkin, Suchir Balaji, Shantanu Jain, William Saunders, Christopher
  Hesse, Andrew~N. Carr, Jan Leike, Joshua Achiam, Vedant Misra, Evan Morikawa,
  Alec Radford, Matthew Knight, Miles Brundage, Mira Murati, Katie Mayer, Peter
  Welinder, Bob McGrew, Dario Amodei, Sam McCandlish, Ilya Sutskever, and
  Wojciech Zaremba. 2021.
\newblock \href {http://arxiv.org/abs/2107.03374} {Evaluating large language
  models trained on code}.
\newblock \emph{CoRR}, abs/2107.03374.

\bibitem[{Cobbe et~al.(2021)Cobbe, Kosaraju, Bavarian, Hilton, Nakano, Hesse,
  and Schulman}]{gsm8k}
Karl Cobbe, Vineet Kosaraju, Mohammad Bavarian, Jacob Hilton, Reiichiro Nakano,
  Christopher Hesse, and John Schulman. 2021.
\newblock \href {http://arxiv.org/abs/2110.14168} {Training verifiers to solve
  math word problems}.
\newblock \emph{CoRR}, abs/2110.14168.

\bibitem[{Dehghani et~al.(2019)Dehghani, Gouws, Vinyals, Uszkoreit, and
  Kaiser}]{universal-tfm}
Mostafa Dehghani, Stephan Gouws, Oriol Vinyals, Jakob Uszkoreit, and Lukasz
  Kaiser. 2019.
\newblock \href {https://openreview.net/forum?id=HyzdRiR9Y7} {Universal
  transformers}.
\newblock In \emph{7th International Conference on Learning Representations,
  {ICLR} 2019, New Orleans, LA, USA, May 6-9, 2019}. OpenReview.net.

\bibitem[{Derezi{\'n}ska and Ha{\l}as(2014)}]{derezinska2014operators}
Anna Derezi{\'n}ska and Konrad Ha{\l}as. 2014.
\newblock Operators for mutation testing of python programs.
\newblock \emph{Res. Rep}.

\bibitem[{Devlin et~al.(2019)Devlin, Chang, Lee, and Toutanova}]{bert}
Jacob Devlin, Ming{-}Wei Chang, Kenton Lee, and Kristina Toutanova. 2019.
\newblock \href {https://doi.org/10.18653/v1/n19-1423} {{BERT:} pre-training of
  deep bidirectional transformers for language understanding}.
\newblock In \emph{Proceedings of the 2019 Conference of the North American
  Chapter of the Association for Computational Linguistics: Human Language
  Technologies, {NAACL-HLT} 2019, Minneapolis, MN, USA, June 2-7, 2019, Volume
  1 (Long and Short Papers)}, pages 4171--4186. Association for Computational
  Linguistics.

\bibitem[{Dong et~al.(2019)Dong, Yang, Wang, Wei, Liu, Wang, Gao, Zhou, and
  Hon}]{unified-lm}
Li~Dong, Nan Yang, Wenhui Wang, Furu Wei, Xiaodong Liu, Yu~Wang, Jianfeng Gao,
  Ming Zhou, and Hsiao{-}Wuen Hon. 2019.
\newblock \href
  {https://proceedings.neurips.cc/paper/2019/hash/c20bb2d9a50d5ac1f713f8b34d9aac5a-Abstract.html}
  {Unified language model pre-training for natural language understanding and
  generation}.
\newblock In \emph{Advances in Neural Information Processing Systems 32: Annual
  Conference on Neural Information Processing Systems 2019, NeurIPS 2019,
  December 8-14, 2019, Vancouver, BC, Canada}, pages 13042--13054.

\bibitem[{Feng et~al.(2020)Feng, Guo, Tang, Duan, Feng, Gong, Shou, Qin, Liu,
  Jiang, and Zhou}]{codebert}
Zhangyin Feng, Daya Guo, Duyu Tang, Nan Duan, Xiaocheng Feng, Ming Gong, Linjun
  Shou, Bing Qin, Ting Liu, Daxin Jiang, and Ming Zhou. 2020.
\newblock \href {https://doi.org/10.18653/v1/2020.findings-emnlp.139}
  {Codebert: {A} pre-trained model for programming and natural languages}.
\newblock In \emph{Findings of the Association for Computational Linguistics:
  {EMNLP} 2020, Online Event, 16-20 November 2020}, volume {EMNLP} 2020 of
  \emph{Findings of {ACL}}, pages 1536--1547. Association for Computational
  Linguistics.

\bibitem[{Graves et~al.(2014)Graves, Wayne, and Danihelka}]{induction1}
Alex Graves, Greg Wayne, and Ivo Danihelka. 2014.
\newblock \href {http://arxiv.org/abs/1410.5401} {Neural turing machines}.
\newblock \emph{CoRR}, abs/1410.5401.

\bibitem[{Graves et~al.(2016)Graves, Wayne, Reynolds, Harley, Danihelka,
  Grabska{-}Barwinska, Colmenarejo, Grefenstette, Ramalho, Agapiou, Badia,
  Hermann, Zwols, Ostrovski, Cain, King, Summerfield, Blunsom, Kavukcuoglu, and
  Hassabis}]{induction4}
Alex Graves, Greg Wayne, Malcolm Reynolds, Tim Harley, Ivo Danihelka, Agnieszka
  Grabska{-}Barwinska, Sergio~Gomez Colmenarejo, Edward Grefenstette, Tiago
  Ramalho, John~P. Agapiou, Adri{\`{a}}~Puigdom{\`{e}}nech Badia, Karl~Moritz
  Hermann, Yori Zwols, Georg Ostrovski, Adam Cain, Helen King, Christopher
  Summerfield, Phil Blunsom, Koray Kavukcuoglu, and Demis Hassabis. 2016.
\newblock \href {https://doi.org/10.1038/nature20101} {Hybrid computing using a
  neural network with dynamic external memory}.
\newblock \emph{Nat.}, 538(7626):471--476.

\bibitem[{Guo et~al.(2022)Guo, Lu, Duan, Wang, Zhou, and Yin}]{unixcoder}
Daya Guo, Shuai Lu, Nan Duan, Yanlin Wang, Ming Zhou, and Jian Yin. 2022.
\newblock \href {https://doi.org/10.18653/v1/2022.acl-long.499} {Unixcoder:
  Unified cross-modal pre-training for code representation}.
\newblock In \emph{Proceedings of the 60th Annual Meeting of the Association
  for Computational Linguistics (Volume 1: Long Papers), {ACL} 2022, Dublin,
  Ireland, May 22-27, 2022}, pages 7212--7225. Association for Computational
  Linguistics.

\bibitem[{Guo et~al.(2021)Guo, Ren, Lu, Feng, Tang, Liu, Zhou, Duan,
  Svyatkovskiy, Fu, Tufano, Deng, Clement, Drain, Sundaresan, Yin, Jiang, and
  Zhou}]{graphcodebert}
Daya Guo, Shuo Ren, Shuai Lu, Zhangyin Feng, Duyu Tang, Shujie Liu, Long Zhou,
  Nan Duan, Alexey Svyatkovskiy, Shengyu Fu, Michele Tufano, Shao~Kun Deng,
  Colin~B. Clement, Dawn Drain, Neel Sundaresan, Jian Yin, Daxin Jiang, and
  Ming Zhou. 2021.
\newblock \href {https://openreview.net/forum?id=jLoC4ez43PZ} {Graphcodebert:
  Pre-training code representations with data flow}.
\newblock In \emph{9th International Conference on Learning Representations,
  {ICLR} 2021, Virtual Event, Austria, May 3-7, 2021}. OpenReview.net.

\bibitem[{Hamlet(1977)}]{mutation-test1}
Richard~G. Hamlet. 1977.
\newblock \href {https://doi.org/10.1109/TSE.1977.231145} {Testing programs
  with the aid of a compiler}.
\newblock \emph{{IEEE} Trans. Software Eng.}, 3(4):279--290.

\bibitem[{Hendrycks et~al.(2021)Hendrycks, Burns, Kadavath, Arora, Basart,
  Tang, Song, and Steinhardt}]{math-dataset}
Dan Hendrycks, Collin Burns, Saurav Kadavath, Akul Arora, Steven Basart, Eric
  Tang, Dawn Song, and Jacob Steinhardt. 2021.
\newblock \href
  {https://datasets-benchmarks-proceedings.neurips.cc/paper/2021/hash/be83ab3ecd0db773eb2dc1b0a17836a1-Abstract-round2.html}
  {Measuring mathematical problem solving with the {MATH} dataset}.
\newblock In \emph{Proceedings of the Neural Information Processing Systems
  Track on Datasets and Benchmarks 1, NeurIPS Datasets and Benchmarks 2021,
  December 2021, virtual}.

\bibitem[{Henighan et~al.(2020)Henighan, Kaplan, Katz, Chen, Hesse, Jackson,
  Jun, Brown, Dhariwal, Gray, Hallacy, Mann, Radford, Ramesh, Ryder, Ziegler,
  Schulman, Amodei, and McCandlish}]{scaling-math}
Tom Henighan, Jared Kaplan, Mor Katz, Mark Chen, Christopher Hesse, Jacob
  Jackson, Heewoo Jun, Tom~B. Brown, Prafulla Dhariwal, Scott Gray, Chris
  Hallacy, Benjamin Mann, Alec Radford, Aditya Ramesh, Nick Ryder, Daniel~M.
  Ziegler, John Schulman, Dario Amodei, and Sam McCandlish. 2020.
\newblock \href {http://arxiv.org/abs/2010.14701} {Scaling laws for
  autoregressive generative modeling}.
\newblock \emph{CoRR}, abs/2010.14701.

\bibitem[{Hindle et~al.(2012)Hindle, Barr, Su, Gabel, and
  Devanbu}]{naturalness}
Abram Hindle, Earl~T. Barr, Zhendong Su, Mark Gabel, and Premkumar~T. Devanbu.
  2012.
\newblock \href {https://doi.org/10.1109/ICSE.2012.6227135} {On the naturalness
  of software}.
\newblock In \emph{34th International Conference on Software Engineering,
  {ICSE} 2012, June 2-9, 2012, Zurich, Switzerland}, pages 837--847. {IEEE}
  Computer Society.

\bibitem[{Jia and Harman(2011)}]{mutation-test2}
Yue Jia and Mark Harman. 2011.
\newblock \href {https://doi.org/10.1109/TSE.2010.62} {An analysis and survey
  of the development of mutation testing}.
\newblock \emph{{IEEE} Trans. Software Eng.}, 37(5):649--678.

\bibitem[{Kaiser and Sutskever(2016)}]{induction3}
Lukasz Kaiser and Ilya Sutskever. 2016.
\newblock \href {http://arxiv.org/abs/1511.08228} {Neural gpus learn
  algorithms}.
\newblock In \emph{4th International Conference on Learning Representations,
  {ICLR} 2016, San Juan, Puerto Rico, May 2-4, 2016, Conference Track
  Proceedings}.

\bibitem[{Kanade et~al.(2020)Kanade, Maniatis, Balakrishnan, and Shi}]{cubert}
Aditya Kanade, Petros Maniatis, Gogul Balakrishnan, and Kensen Shi. 2020.
\newblock \href {http://proceedings.mlr.press/v119/kanade20a.html} {Learning
  and evaluating contextual embedding of source code}.
\newblock In \emph{Proceedings of the 37th International Conference on Machine
  Learning, {ICML} 2020, 13-18 July 2020, Virtual Event}, volume 119 of
  \emph{Proceedings of Machine Learning Research}, pages 5110--5121. {PMLR}.

\bibitem[{Kurach et~al.(2016)Kurach, Andrychowicz, and Sutskever}]{induction2}
Karol Kurach, Marcin Andrychowicz, and Ilya Sutskever. 2016.
\newblock \href {http://arxiv.org/abs/1511.06392} {Neural random-access
  machines}.
\newblock In \emph{4th International Conference on Learning Representations,
  {ICLR} 2016, San Juan, Puerto Rico, May 2-4, 2016, Conference Track
  Proceedings}.

\bibitem[{Ling et~al.(2017)Ling, Yogatama, Dyer, and Blunsom}]{aqua}
Wang Ling, Dani Yogatama, Chris Dyer, and Phil Blunsom. 2017.
\newblock \href {https://doi.org/10.18653/v1/P17-1015} {Program induction by
  rationale generation: Learning to solve and explain algebraic word problems}.
\newblock In \emph{Proceedings of the 55th Annual Meeting of the Association
  for Computational Linguistics, {ACL} 2017, Vancouver, Canada, July 30 -
  August 4, Volume 1: Long Papers}, pages 158--167. Association for
  Computational Linguistics.

\bibitem[{Lu et~al.(2022)Lu, Grover, Abbeel, and Mordatch}]{frozen-tfm}
Kevin Lu, Aditya Grover, Pieter Abbeel, and Igor Mordatch. 2022.
\newblock \href {https://ojs.aaai.org/index.php/AAAI/article/view/20729}
  {Frozen pretrained transformers as universal computation engines}.
\newblock In \emph{Thirty-Sixth {AAAI} Conference on Artificial Intelligence,
  {AAAI} 2022, Thirty-Fourth Conference on Innovative Applications of
  Artificial Intelligence, {IAAI} 2022, The Twelveth Symposium on Educational
  Advances in Artificial Intelligence, {EAAI} 2022 Virtual Event, February 22 -
  March 1, 2022}, pages 7628--7636. {AAAI} Press.

\bibitem[{Nye et~al.(2021)Nye, Andreassen, Gur{-}Ari, Michalewski, Austin,
  Bieber, Dohan, Lewkowycz, Bosma, Luan, Sutton, and Odena}]{scratchpad}
Maxwell~I. Nye, Anders~Johan Andreassen, Guy Gur{-}Ari, Henryk Michalewski,
  Jacob Austin, David Bieber, David Dohan, Aitor Lewkowycz, Maarten Bosma,
  David Luan, Charles Sutton, and Augustus Odena. 2021.
\newblock \href {http://arxiv.org/abs/2112.00114} {Show your work: Scratchpads
  for intermediate computation with language models}.
\newblock \emph{CoRR}, abs/2112.00114.

\bibitem[{Puri et~al.(2021)Puri, Kung, Janssen, Zhang, Domeniconi, Zolotov,
  Dolby, Chen, Choudhury, Decker, Thost, Buratti, Pujar, Ramji, Finkler,
  Malaika, and Reiss}]{codenet}
Ruchir Puri, David~S. Kung, Geert Janssen, Wei Zhang, Giacomo Domeniconi,
  Vladimir Zolotov, Julian Dolby, Jie Chen, Mihir~R. Choudhury, Lindsey Decker,
  Veronika Thost, Luca Buratti, Saurabh Pujar, Shyam Ramji, Ulrich Finkler,
  Susan Malaika, and Frederick Reiss. 2021.
\newblock \href
  {https://datasets-benchmarks-proceedings.neurips.cc/paper/2021/hash/a5bfc9e07964f8dddeb95fc584cd965d-Abstract-round2.html}
  {Codenet: {A} large-scale {AI} for code dataset for learning a diversity of
  coding tasks}.
\newblock In \emph{Proceedings of the Neural Information Processing Systems
  Track on Datasets and Benchmarks 1, NeurIPS Datasets and Benchmarks 2021,
  December 2021, virtual}.

\bibitem[{Radford et~al.(2018)Radford, Narasimhan, Salimans, Sutskever
  et~al.}]{gpt}
Alec Radford, Karthik Narasimhan, Tim Salimans, Ilya Sutskever, et~al. 2018.
\newblock \href
  {https://s3-us-west-2.amazonaws.com/openai-assets/research-covers/language-unsupervised/language_understanding_paper.pdf}
  {Improving language understanding by generative pre-training}.

\bibitem[{Radford et~al.(2019)Radford, Wu, Child, Luan, Amodei, Sutskever
  et~al.}]{gpt2}
Alec Radford, Jeffrey Wu, Rewon Child, David Luan, Dario Amodei, Ilya
  Sutskever, et~al. 2019.
\newblock Language models are unsupervised multitask learners.
\newblock \emph{OpenAI blog}, 1(8):9.

\bibitem[{Raffel et~al.(2020)Raffel, Shazeer, Roberts, Lee, Narang, Matena,
  Zhou, Li, and Liu}]{t5}
Colin Raffel, Noam Shazeer, Adam Roberts, Katherine Lee, Sharan Narang, Michael
  Matena, Yanqi Zhou, Wei Li, and Peter~J. Liu. 2020.
\newblock \href {http://jmlr.org/papers/v21/20-074.html} {Exploring the limits
  of transfer learning with a unified text-to-text transformer}.
\newblock \emph{J. Mach. Learn. Res.}, 21:140:1--140:67.

\bibitem[{Reed and de~Freitas(2016)}]{induction5}
Scott~E. Reed and Nando de~Freitas. 2016.
\newblock \href {http://arxiv.org/abs/1511.06279} {Neural
  programmer-interpreters}.
\newblock In \emph{4th International Conference on Learning Representations,
  {ICLR} 2016, San Juan, Puerto Rico, May 2-4, 2016, Conference Track
  Proceedings}.

\bibitem[{Saxton et~al.(2019)Saxton, Grefenstette, Hill, and
  Kohli}]{deepmind-math}
David Saxton, Edward Grefenstette, Felix Hill, and Pushmeet Kohli. 2019.
\newblock \href {https://openreview.net/forum?id=H1gR5iR5FX} {Analysing
  mathematical reasoning abilities of neural models}.
\newblock In \emph{7th International Conference on Learning Representations,
  {ICLR} 2019, New Orleans, LA, USA, May 6-9, 2019}. OpenReview.net.

\bibitem[{Svyatkovskiy et~al.(2020)Svyatkovskiy, Deng, Fu, and
  Sundaresan}]{gpt-c}
Alexey Svyatkovskiy, Shao~Kun Deng, Shengyu Fu, and Neel Sundaresan. 2020.
\newblock \href {https://doi.org/10.1145/3368089.3417058} {Intellicode compose:
  code generation using transformer}.
\newblock In \emph{{ESEC/FSE} '20: 28th {ACM} Joint European Software
  Engineering Conference and Symposium on the Foundations of Software
  Engineering, Virtual Event, USA, November 8-13, 2020}, pages 1433--1443.
  {ACM}.

\bibitem[{Tay et~al.(2021)Tay, Dehghani, Abnar, Shen, Bahri, Pham, Rao, Yang,
  Ruder, and Metzler}]{long-range}
Yi~Tay, Mostafa Dehghani, Samira Abnar, Yikang Shen, Dara Bahri, Philip Pham,
  Jinfeng Rao, Liu Yang, Sebastian Ruder, and Donald Metzler. 2021.
\newblock \href {https://openreview.net/forum?id=qVyeW-grC2k} {Long range arena
  : {A} benchmark for efficient transformers}.
\newblock In \emph{9th International Conference on Learning Representations,
  {ICLR} 2021, Virtual Event, Austria, May 3-7, 2021}. OpenReview.net.

\bibitem[{Tay et~al.(2022)Tay, Dehghani, Bahri, and Metzler}]{tay2022efficient}
Yi~Tay, Mostafa Dehghani, Dara Bahri, and Donald Metzler. 2022.
\newblock \href {https://dl.acm.org/doi/full/10.1145/3530811} {Efficient
  transformers: A survey}.
\newblock \emph{ACM Computing Surveys}, 55(6):1--28.

\bibitem[{Vaswani et~al.(2017)Vaswani, Shazeer, Parmar, Uszkoreit, Jones,
  Gomez, Kaiser, and Polosukhin}]{vaswani2017attention}
Ashish Vaswani, Noam Shazeer, Niki Parmar, Jakob Uszkoreit, Llion Jones,
  Aidan~N Gomez, {\L}ukasz Kaiser, and Illia Polosukhin. 2017.
\newblock Attention is all you need.
\newblock \emph{Advances in neural information processing systems}, 30.

\bibitem[{Velickovic et~al.(2020{\natexlab{a}})Velickovic, Buesing, Overlan,
  Pascanu, Vinyals, and Blundell}]{induction6}
Petar Velickovic, Lars Buesing, Matthew~C. Overlan, Razvan Pascanu, Oriol
  Vinyals, and Charles Blundell. 2020{\natexlab{a}}.
\newblock \href
  {https://proceedings.neurips.cc/paper/2020/hash/176bf6219855a6eb1f3a30903e34b6fb-Abstract.html}
  {Pointer graph networks}.
\newblock In \emph{Advances in Neural Information Processing Systems 33: Annual
  Conference on Neural Information Processing Systems 2020, NeurIPS 2020,
  December 6-12, 2020, virtual}.

\bibitem[{Velickovic et~al.(2020{\natexlab{b}})Velickovic, Ying, Padovano,
  Hadsell, and Blundell}]{induction7}
Petar Velickovic, Rex Ying, Matilde Padovano, Raia Hadsell, and Charles
  Blundell. 2020{\natexlab{b}}.
\newblock \href {https://openreview.net/forum?id=SkgKO0EtvS} {Neural execution
  of graph algorithms}.
\newblock In \emph{8th International Conference on Learning Representations,
  {ICLR} 2020, Addis Ababa, Ethiopia, April 26-30, 2020}. OpenReview.net.

\bibitem[{Wang et~al.(2021{\natexlab{a}})Wang, Wang, Mi, Zhou, Wan, Liu, Li,
  Wu, Liu, and Jiang}]{syncobert}
Xin Wang, Yasheng Wang, Fei Mi, Pingyi Zhou, Yao Wan, Xiao Liu, Li~Li, Hao Wu,
  Jin Liu, and Xin Jiang. 2021{\natexlab{a}}.
\newblock Syncobert: Syntax-guided multi-modal contrastive pre-training for
  code representation.
\newblock \emph{arXiv preprint arXiv:2108.04556}.

\bibitem[{Wang et~al.(2020)Wang, Wang, Gao, and Wang}]{ginn}
Yu~Wang, Ke~Wang, Fengjuan Gao, and Linzhang Wang. 2020.
\newblock \href {https://doi.org/10.1145/3428205} {Learning semantic program
  embeddings with graph interval neural network}.
\newblock \emph{Proc. {ACM} Program. Lang.}, 4({OOPSLA}):137:1--137:27.

\bibitem[{Wang et~al.(2021{\natexlab{b}})Wang, Wang, Joty, and Hoi}]{codet5}
Yue Wang, Weishi Wang, Shafiq~R. Joty, and Steven C.~H. Hoi.
  2021{\natexlab{b}}.
\newblock \href {https://doi.org/10.18653/v1/2021.emnlp-main.685} {Codet5:
  Identifier-aware unified pre-trained encoder-decoder models for code
  understanding and generation}.
\newblock In \emph{Proceedings of the 2021 Conference on Empirical Methods in
  Natural Language Processing, {EMNLP} 2021, Virtual Event / Punta Cana,
  Dominican Republic, 7-11 November, 2021}, pages 8696--8708. Association for
  Computational Linguistics.

\bibitem[{Yan et~al.(2020)Yan, Swersky, Koutra, Ranganathan, and
  Hashemi}]{Neural-Execution-Engines}
Yujun Yan, Kevin Swersky, Danai Koutra, Parthasarathy Ranganathan, and Milad
  Hashemi. 2020.
\newblock \href
  {https://proceedings.neurips.cc/paper/2020/hash/c8b9abffb45bf79a630fb613dcd23449-Abstract.html}
  {Neural execution engines: Learning to execute subroutines}.
\newblock In \emph{Advances in Neural Information Processing Systems 33: Annual
  Conference on Neural Information Processing Systems 2020, NeurIPS 2020,
  December 6-12, 2020, virtual}.

\bibitem[{Zaremba and Sutskever(2014)}]{learn-to-execute}
Wojciech Zaremba and Ilya Sutskever. 2014.
\newblock \href {http://arxiv.org/abs/1410.4615} {Learning to execute}.
\newblock \emph{CoRR}, abs/1410.4615.

\bibitem[{Zhou et~al.(2022)Zhou, Nova, Larochelle, Courville, Neyshabur, and
  Sedghi}]{algo-reason}
Hattie Zhou, Azade Nova, Hugo Larochelle, Aaron~C. Courville, Behnam Neyshabur,
  and Hanie Sedghi. 2022.
\newblock \href {https://doi.org/10.48550/arXiv.2211.09066} {Teaching
  algorithmic reasoning via in-context learning}.
\newblock \emph{CoRR}, abs/2211.09066.

\end{thebibliography}
\bibliographystyle{acl_natbib}

\appendix
\section{Dataset Detail}
\label{sec:appendix-dataset}
To obtain executable programs, we build the Python Code Execution dataset based on submissions to competitive programming problems from CodeNet \cite{codenet}. 
These human-written programs with real-world complexity are derived from online judge websites AIZU \footnote{\url{https://onlinejudge.u-aizu.ac.jp/}} and AtCoder \footnote{\url{https://atcoder.jp/}}.
CodeNet contains 240k Python submissions, aiming to solve 8,00 distinct programming problems. 
Each submission is a single-file Python program that
reads from stdin and writes to stdout. 
Each programming problem provides at least one sample input and at most four sample inputs.
Since executing a program relies on an input, we replace the statements that read from input streams with assignment statements that assign input values to variables. 
We run each submission in a sandbox environment to get the execution trace for that program.
Programs are restricted to one second of execution time and 1024 lines of execution trace, and will be filtered out if they exceed the limits.
We also remove the programs that result in runtime errors during parsing or execution, by catching Python exceptions raised in programs.
This results in a dataset of 387k executable programs, each paired with a trace. 

To construct a large-scale dataset of executable programs, 
we propose a mutation-based data augmentation approach.
we first present a set of mutation operators as shown in Table \ref{mutation-operators}.
Most of them correspond to selected operators used in strongly typed general purpose languages and are adopted to the Python language. 
Operators designed for Python features are also included, such as Slice Index Removal (SIR) and Reverse Iteration Loop (RIL). 
Then we leverage the tree-sitter\footnote{\url{https://tree-sitter.github.io/tree-sitter/}} to 
convert a program into an abstract syntax tree and then extract its node type information to get a candidate list of all mutable literals, operators and statements. 
For each mutable candidate, we apply the related mutation operators with 50\% probability. 
Specifically, we change a numeric literal $x$ into a random number from a Gaussian distribution with mean $x$ and standard deviation 100. We either extend a string with one or two random characters or shorten a string. We randomly pick one of the three loop-related operators or keep it as it is when handling each loop.
All operators can be applied before a mutated program execution, and possible mutants with errors are to be detected and eliminated during execution. 
By mutating each program 20 times, we obtain 3.2M deduplicated programs, each paired with a trace.

We use the CodeNet Mutants (CodeNetMut) to build the pre-training dataset. 
To prevent data leakage, all submissions to the same problem become part of the same split. 
We use submissions of 710 problems with their mutants to build the pre-training dataset. 
Since mutation greatly enhances diversity, these programs embody rich semantics and complex operations.
Other submissions (without mutations) are used to build the validation and test dataset. These human-authored programs ensure the quality of evaluation data.

\begin{table*}
\centering
\small
\begin{tabular}{lccc}
\toprule
\textbf{Model} & \textbf{Stage1 (S1)} & \textbf{Stage2 (S2)} & \textbf{Stage3 (S3)} \\
\midrule 
CEL & SingleLine & Tutorial & CodeNetMut \\
CodeExecutor &  SingleLine & SingleLine (3M), Tutorial & SingleLine (3M), Tutorial, CodeNetMut \\
\bottomrule
\end{tabular}
\caption{\label{three-stages}
Datasets that CEL and CodeExecutor use for three-stage pre-training. ``SingleLine (3M)'' denotes 3 million instances within SingleLine that are most difficult for CodeExecutor to generate. 
}
\end{table*}

\begin{figure*}[t]
  \centering  
  \includegraphics[width=0.98\linewidth]
  {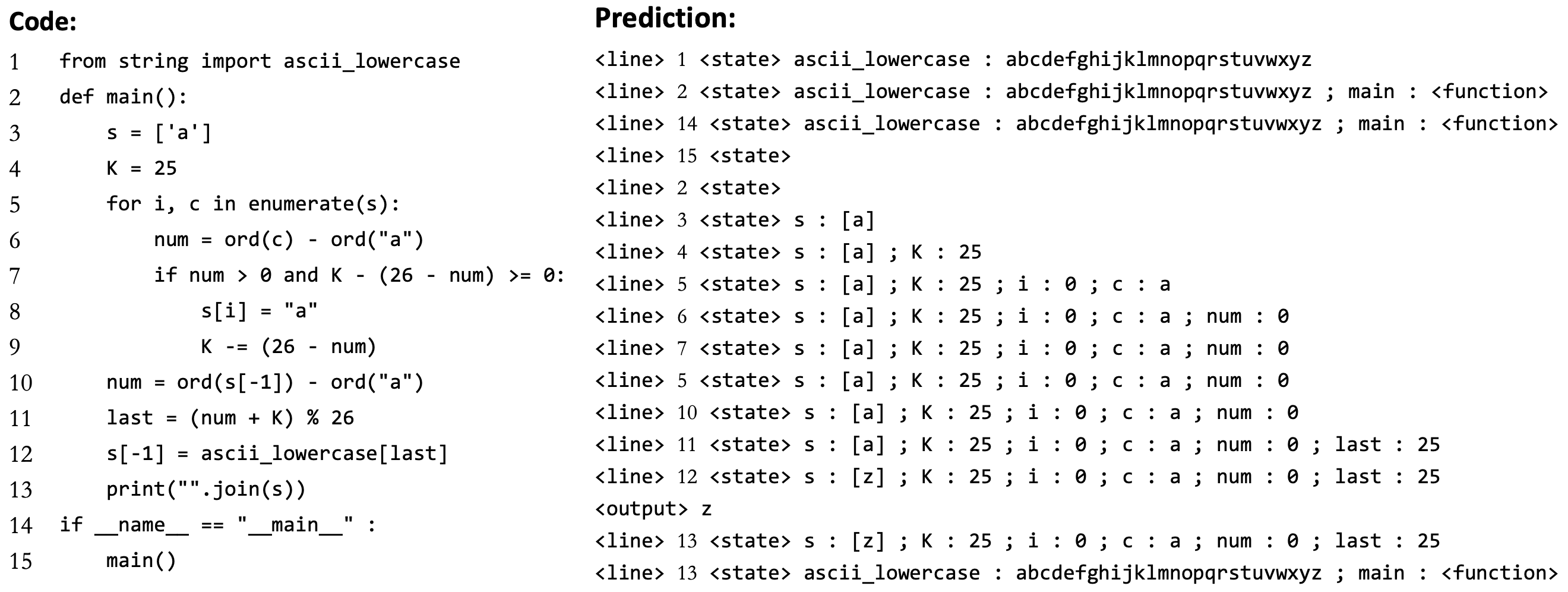} 
  \caption{An example from CodeNetMut test split, which covers all the categories of Python programming knowledge. CodeExecutor gives the correct prediction.}
  \label{fig:example-c}
\end{figure*}

\section{Model Configurations} 
\label{sec:appendix-model}
We build our model based on 12 layers of Transformer with 768 dimensional hidden states and 12 attention heads. 
We add 210 additional special tokens into the vocabulary to represent 200 line numbers, 3 pre-training dataset names, and trace structure described in §\ref{sssec:4.2}.
During pre-training, we set the max length of input sequence and batch size to be 1024 and 256, respectively. 
We use the Adam optimizer to update model parameters with 4e-4 learning rate. 
We first employ the SingleLine dataset to pre-train the model with the code execution objective for 500k steps.
We then reserve 3 million instances in SingleLine that are most difficult for our model to generate and add Tutorial data into the corpus, pre-training for 300k steps.
We add CodeNetMut into the corpus and further pre-train for 300k steps.
We pre-train the model on a cluster of 16 NVIDIA Tesla V100 with 32GB memory and the total training time is about a month.
For inference, we set beam search as 10.

\section{Three-stage Pre-training}
\label{appedix:stages}
In table \ref{three-stages}, we list the datasets that CodeExecutorLimited (CEL) and CodeExecutor use for three-stage pre-training, respectively. 

The first stage of pre-training for CEL uses the SingleLine dataset, resulting in the model CEL-S1. 
In the second stage, CEL is initialized with CEL-S1 and pre-trained with the Tutorial dataset, resulting in the model CEL-S2. 
In the third stage, CEL is initialized with CEL-S2 and pre-trained with the CodeNetMut dataset, resulting in the model CEL-S3. 

On the other hand, CodeExecutor is first pre-trained with the SingleLine dataset, then the 3 million most challenging SingleLine data is selected for later training stages based on the model's loss. 
In the second stage, CodeExecutor is pre-trained with the 3 million difficult SingleLine data, along with the Tutorial dataset. In the third stage, CodeExecutor is pre-trained with the 3 million difficult SingleLine data, the entire Tutorial dataset, and the CodeNetMut dataset.

\section{Qualitative Examples} 
\label{appedix:examples}
Additional examples are shown here.

Figure \ref{fig:example-c} shows an example that covers all the categories of Python programming knowledge in Table \ref{human-evaluation}. CodeExecutor generates the same trace as ground truth.

\begin{figure}[t]
  \centering  
  \includegraphics[width=0.7\linewidth]
  {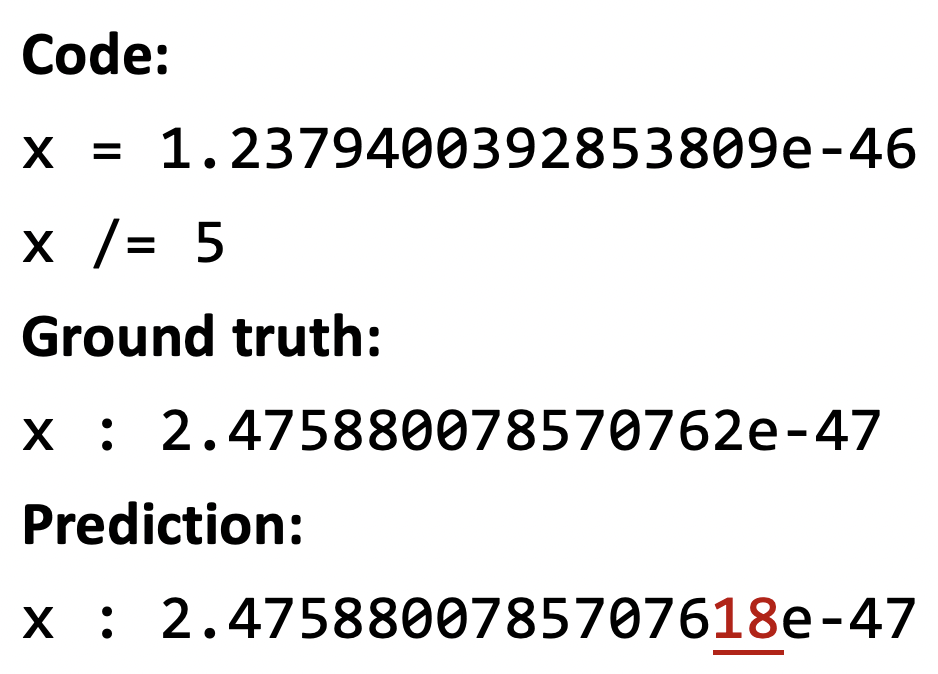} 
  \caption{An example of division calculations with decimals, where CodeExecutor correctly produce the first fifteen digits, with mistakes highlighted by an underline.}
  \label{fig:example-div}
\end{figure}

Figure \ref{fig:example-div} is an example of performing division calculations with decimals.
CodeExecutor is able to produce the correct first fifteen digits and makes errors in the remaining two digits.

\end{document}